\documentclass{elsart}
\usepackage{amsmath,amssymb,graphics,epsfig,mcite}

\newcommand{\Tr}[0]{\text{Tr\;}}
\newcommand{\bra}[1]{\left\langle #1\right |}
\newcommand{\ket}[1]{\left| #1\right \rangle}

\begin{document}
\begin{frontmatter}
\title{New solutions to the Ginsparg-Wilson equation}
\author[a]{Nigel Cundy}
\address[a]{Institut f\"ur Theoretische Physik, Universit\"at Regensburg, D-93040 Regensburg, Germany}

\begin{abstract}
The overlap operator is just the simplest of a class of Dirac operators with an exact chiral symmetry. I demonstrate how a general class of chiral Dirac operators can be constructed, show that they have no fermion doublers and that they are all exponentially local, and test my conclusions numerically for a few examples. However, since these operators are more expensive than the overlap operator, it is unlikely that they will be useful in practical simulations. 
\end{abstract}
\end{frontmatter}
\section{Introduction}\label{sec:1}
For many years, it seemed that simulating chiral symmetry on the lattice was impossible, because of the Nielson-Ninomiya theorem~\cite{Nishy-Ninny}, which, in one form, states that it is impossible to have a Dirac operator which is local, has the correct continuum limit, hyper-cubic symmetry, no doublers, and anti-commutes with $\gamma_5$. A method of avoiding the Nielson-Ninomiya theorem was suggested shortly afterwards, in 1982, by Ginsparg and Wilson~\cite{Ginsparg:1982bj}, who proposed that the smoothest way of breaking chiral symmetry on the lattice in the continuum limit  is to use a Dirac operator which, rather than anti-commuting with $\gamma_5$, satisfies the Ginsparg-Wilson relation, 
\begin{gather}  
aD\gamma_5+a\gamma_5D = a^2\frac{1}{2}D(\gamma_5R+R\gamma_5) D.\label{eq:GW1}
\end{gather}
I will call $R$ the Ginsparg-Wilson function, and it plays a key role in this paper. 

 The Ginsparg-Wilson relation was soon forgotten as no solutions were known. After this relation was rediscovered in 1998~\cite{Hasenfratz:1998ri}, Martin L\"uscher showed that it implied an exact symmetry of the QCD Lagrangian for any lattice Dirac operator obeying the Ginsparg-Wilson relation with a local $R$, and that this symmetry reduces to  chiral symmetry in the continuum limit~\cite{Luscher:1998pqa}.  Inspired by the work of Kaplan~\cite{Kaplan:1992bt} and using an infinite number of fermion fields to avoid the Nielson-Ninomiya theorem, Neuberger and Narayanan had already, a few years earlier, found a Dirac operator which satisfied the Ginsparg-Wilson relation with $R = 1$, namely the overlap operator~\cite{Narayanan:1993sk,*Narayanan:1993ss}, which can be written in the form given in section \ref{sec:2}, equations (\ref{eq:overlap_operator}) and (\ref{eq:DW}). Non-local solutions to the Ginsparg-Wilson equation have also been found, for example in ~\cite{deA.Bicudo:1999wq}, and the various fixed point actions, including the classically perfect action,~\cite{Hasenfratz:1994,*Bietenholz:1995nk} are known to satisfy the Ginsparg Wilson relation~\cite{Hasenfratz:1998ri}, but in practice a truncated form of the Dirac operator has to be used, meaning that the chiral symmetry becomes inexact. Fujikawa has constructed the algebraic solutions to a particular form of a generalised Ginsparg-Wilson equation~\cite{Fujikawa:2000my,*Fujikawa:2001fb}. The construction of these operators requires two steps: firstly constructing an intermediate operator from the matrix sign function of a Wilson-type operator, then taking a root of that operator. Thus, these operators effectively require a nested series of roots of a matrix, meaning that they will be considerably slower than overlap fermions in practice. Werner Kerler has constructed generalised chiral lattice Dirac operators by solving a particular form of the Ginsparg Wilson relation~\cite{Kerler:2002xk,*Kerler:2002fi,*Kerler:2003qv}, and I discuss the relationship between his work and my own in section \ref{sec:5}. The overlap operator, Fujikawa's and Kerler's solutions are all particular forms of my most general solution. I shall not discuss the generalisation of the Ginsparg-Wilson equation obtained by using different kernels in the matrix sign function.

Of course, not every possible Ginsparg-Wilson function $R$ will lead to a lattice chiral symmetry. An easy, though over simplified, way of seeing this is to note that if, in an expansion in the lattice spacing, $R$ is inversely proportional to the lattice spacing, $a$, (or worse) then the right hand side of equation (\ref{eq:GW1}) will remain constant (or diverge) rather than reduce smoothly to zero in the continuum limit. More strictly, the lattice chiral symmetry requires that $\{\gamma_5,R\}$ is non-vanishing and local~\cite{Hasenfratz:1998jp}. This places a number of restrictions on the lattice Dirac operator $D$, such as that if it satisfies the Ginsparg-Wilson symmetry and  has no doublers then it cannot be ultra-local~\cite{Horvath:1999bk}, but at best, like the overlap operator~\cite{Hernandez:1998et}, exponentially local. 


I define $\mathfrak{D}_C$ as the class of possible suitable Dirac operators (local, $\gamma_5$-Hermitian, with the correct continuum limit, and no doublers) which have their eigenvalues lying on a curve in the complex plane. It is interesting that the overlap operator, the generalisations by Fujikawa and Kerler, and the Dirac operator in the continuum are all within $\mathfrak{D}_C$. In the continuum, the eigenvalue spectrum is a straight line along the imaginary axis; for overlap fermions, the curve is a circle. However, for example, Wilson fermions, which have no chiral symmetry, and staggered fermions, which have doublers, are outside $\mathfrak{D}_C$. It is an interesting hypothesis whether all suitable lattice Dirac operators within $\mathfrak{D}_C$ satisfy Ginsparg-Wilson chiral symmetry. In this paper, I demonstrate the plausibility of this hypothesis by showing that a class of Dirac operators with the correct continuum limit, no fermion doublers, and with their eigenvalue spectrum lying on an arbitrary curve in the complex plane are all suitable chiral fermions --- in that they obey a Ginsparg-Wilson relation with local $R$ and are local themselves.  

It is also worth considering the reverse hypothesis, whether all possible Ginsparg-Wilson lattice Dirac operators are within the class $\mathfrak{D}_C$. In section \ref{sec:comment}, I demonstrate that this is true for Dirac operators satisfying $[D,D^{\dagger}] = 0$, a condition obeyed in the continuum. However, in general it does not seem to be the case. Certain fixed point lattice Dirac operators~\cite{Hasenfratz:1998ri,BietenholzPersonalCommunication} offer one known counterexample, and I discuss others in section \ref{sec:comment}. However, these Dirac operators are related to a Dirac operator within $\mathfrak{D}_C$ by a simple chirally invariant transformation. However, there are distinct homotopy classes within $\mathfrak{D}_C$ which cannot be mapped to each other by this transformation.

In section \ref{sec:2}, I review the Ginsparg-Wilson chiral symmetry, to give the notation and tools which I will use in section \ref{sec:3} to construct the new Ginsparg Wilson operators and demonstrate that, with sufficiently smooth gauge fields, they are exponentially local. I test the locality numerically for a few examples in section \ref{sec:4} and, after outlining some generalisations to this work in section \ref{sec:5}, I conclude in section \ref{sec:conclusion}.

\section{Ginsparg-Wilson chiral symmetry}\label{sec:2}
\subsection{Introduction}
In this section, I review the demonstration that the Ginsparg-Wilson equation implies an exact lattice chiral symmetry and gives a lattice topological charge satisfying an index theorem. This section is based on the work of Martin L\"uscher~\cite{Luscher:1998pqa}.

Consider the following ``chiral" transformation of the fermion fields:
\begin{gather}
\psi_i' = e^{\alpha \gamma_5 \left(S - \frac{1}{2}aRD\right)}\psi_i\nonumber\\
\overline{\psi}'_i = \overline{\psi}_ie^{\alpha \left(S - \frac{1}{2}aDR\right)\gamma_5},
\end{gather}
which for small enough $\alpha$ can be written as
\begin{gather}
\psi'_i = \psi_i+\alpha \gamma_5 \left(S - \frac{1}{2}aRD\right)\psi_i\nonumber\\
\overline{\psi}'_i=\overline{\psi}_i+ \overline{\psi}_i\alpha \left(S - \frac{1}{2}aDR\right)\gamma_5.
\end{gather}
If there is  a mass-less fermionic action
\begin{gather}
S_f = \sum_{i =1}^{N_f} \overline{\psi}_iD\psi_i,
\end{gather}
where $D$ is $\gamma_5$-Hermitian ($D^{\dagger} = \gamma_5 D \gamma_5$), then it is trivial to demonstrate that the action is conserved under the chiral transformation if and only if the general Ginsparg Wilson relation is fulfilled
\begin{gather}
D \gamma_5 S + S \gamma_5 D = a\frac{1}{2} D (R \gamma_5 + \gamma_5 R) D.\label{eq:5}
\end{gather}
We can write $R = R_A + R_C$, where $R_A = \frac{1}{2}(R-\gamma_5R\gamma_5)$ and $R_C = \frac{1}{2}(R+\gamma_5R\gamma_5)$ so that $\{R_A,\gamma_5\}=0$ and  $[R_C,\gamma_5]=0$. Since $R_A$ does not contribute to the Ginsparg-Wilson relation, we can simplify without any loss of generalisation by restricting $R$ to those functions which commute with $\gamma_5$. Additionally I restrict $S$ to $\gamma_5$-Hermitian operators.  By taking the Hermitian conjugate of equation (\ref{eq:5}), it is clear that, with these restrictions, $R$ must be Hermitian. 
It is now possible to simplify equation (\ref{eq:5}) so that it reads
\begin{gather}
D^{\dagger} S + S^{\dagger}D = aD^{\dagger}RD.\label{eq:GW}
\end{gather} 
This can be associated with the continuum chiral symmetry because in the continuum limit (for suitable $S$ and $R$, i.e. they are local and with the correct form in the continuum) it reduces to $\gamma_5 D + D \gamma_5 = 0$. The Ginsparg-Wilson relation is usually defined with $S=1$, and $S$ can easily be absorbed into the definition of $D$ and $R$.
Of course, the Dirac operator must also satisfy the usual criteria for a suitable Dirac operator, namely that it is exponentially local (according to the definition $D_{xy} \le \alpha e^{-\beta|x-y|}$ for positive $\alpha$ and $\beta$), that the Dirac operator has no fermion doublers, and that expanding the Fourier transformed operators in the lattice spacing, $a$, gives
\begin{align}
\tilde{D}(p) =& ia\gamma_{\mu} (p_{\mu} + A^b_{\mu}T^b) + O(a^2)\nonumber\\ 
\tilde{S}(p) = & 1 + O(a)\nonumber\\
\tilde{R}(p) = & O(1),
\end{align}
where $T^b$ are the (Hermitian) generators of the gauge group, and $A_{\mu} = A^b_{\mu}T^b$ represents the gauge fields.
It can be shown that a Ginsparg-Wilson Dirac operator correctly resolves the U(1) anomaly by noting that the fermion measure is not invariant under this transformation: 
\begin{gather}
d\psi' d\overline{\psi}' = \det\left| e^{\alpha \gamma_5 \left(S - \frac{1}{2}aRD\right)} \right|\det\left|e^{\alpha \left(S - \frac{1}{2}aDR\right)\gamma_5}\right|d\psi d\overline{\psi},
\end{gather} 
which leads to a definition of a lattice topological index~\cite{Luscher:1998pqa,Kikukawa:1998pd,*Adams:1998eg}
\begin{gather}
Q_f = \Tr(\gamma_5 S - \frac{1}{2}a\gamma_5RD).
\end{gather}
As an aside, I note that, in the continuum limit, the Yang Mills action can be written as~\cite{Horvath:2006md} 
\begin{gather}
\frac{1}{4}F_{\mu\nu}^2 = c_s\Tr(S - \frac{1}{2} a RD) + constant,
\end{gather}
and the electromagnetic field tensor as~\cite{Liu:2007hq}
\begin{gather}
F_{\mu\nu} = c_f\Tr\;\sigma_{\mu\nu}(S - \frac{1}{2} a RD),
\end{gather}
where $c_s$ and $c_f$ are normalisation constants. This means that all the elements of QCD and the electro-weak Lagrangians can be constructed from any Ginsparg-Wilson operator.

$R$ can, of course, be trivially constructed algebraically for any lattice Dirac operator (assuming that its inverse exists),
\begin{gather}
R = S \frac{1}{D} + \frac{1}{D^{\dagger}}S^{\dagger},
\end{gather}
but only a few possible Dirac operators will give the local $R$ needed for the lattice chiral symmetry. It is trivial to show that inserting the Wilson Dirac operator gives a non-local $R$ since the inverse of the Wilson operator is non-local and has additive mass renormalisation~\cite{Bietenholz:1999dh}.  

One exact solution to the Ginsparg-Wilson equation, with local $R$, has been known for over ten years. The mass-less overlap operator~\cite{Neuberger:1998my}, given by
\begin{gather}
D_O = 1 + \gamma_5 \epsilon(\gamma_5 D_W),\label{eq:overlap_operator}
\end{gather}
where $\epsilon$ is the matrix sign function, satisfies the Ginsparg-Wilson equation with $R=1$.
$D_W$ can in principle be any valid lattice Dirac operator with the correct continuum limit, no fermion doublers, and a suitably chosen mass term (e.g. in equation (\ref{eq:DW}), any $m$ satisfying $m_c<m<2$, where $m_c\sim 0$ is the Wilson critical mass, will suffice, although in practice $m$ should be tuned to improve the locality of the operator). For the purposes of this work, I will use the simplest possibility, the Wilson Dirac operator, which I shall write as
\begin{gather}
D_W = \frac{1}{2}\sum_{\mu}\left[\gamma_{\mu}(\partial_{\mu} + \partial^{*}_{\mu}) - \partial^{*}_{\mu}\partial_{\mu}\right] - m.\label{eq:DW}
\end{gather}
$\partial_{\mu}$ and $\partial^*_{\mu}$ are the forward and backward lattice Dirac operators, defined as
\begin{gather}
\partial_{\mu}(\psi(x)) = e^{iaA_{\mu}(x + a\frac{\hat{\mu}}{2})}\psi(x + a\hat{\mu}) - \psi(x)\nonumber\\
\partial^*_{\mu}(\psi(x)) = \psi(x ) - e^{-iaA_{\mu}(x - a\frac{\hat{\mu}}{2})}\psi(x-a\hat{\mu}).
\end{gather} 
Thus, taking the Fourier transform to obtain the momentum representation of $D_W$, I obtain
\begin{align}
\tilde{D}_W(p) =& i\gamma_{\mu}H_{\mu} + W -m\nonumber\\
H_{\mu} = & -\frac{i}{2}\sum_x e^{i(p,x)} \left(e^{iaA_{\mu}(x+a\frac{\hat{\mu}}{2})}e^{i a p_{\mu}} - e^{-iaA_{\mu}(x-a\frac{\hat{\mu}}{2})}e^{-i a p_{\mu}}\right)\nonumber\\
W = &-\frac{1}{2}\sum_{\mu,x,y}e^{i(p,x)}e^{i(p,y)}\left(e^{iaA_{\mu}(x+a\frac{\hat{\mu}}{2})}e^{i a p_{\mu}} + e^{-iaA_{\mu}(y-a\frac{\hat{\mu}}{2})}e^{-i a p_{\mu}} \right.\nonumber\\&\phantom{some copious amounts of space}\left.- 1 - e^{iaA_{\mu}(x+a\frac{\hat{\mu}}{2})}e^{-iaA_{\mu}(y-a\frac{\hat{\mu}}{2})}\right)
.\label{eq:dwp}
\end{align}
Note that in the free theory,
\begin{align}
H_{\mu} =& \sin(ap_{\mu})\nonumber\\
W =& \sum_{\mu} 2 \sin^2\frac{a p_{\mu}}{2}.
\end{align}
Using equation (\ref{eq:dwp}), the momentum representation of the overlap operator can be expressed as
\begin{align}
\tilde{D}_O(p) =& 1 + (\gamma_{\mu}H_{\mu} + W - m)\frac{1}{\sqrt{\tilde{B}(p)}}\nonumber\\
\tilde{B}(p) =& (m-W)^2 + H_{\mu}H_{\mu} + i \gamma_{\mu}\left[(W-m)H_{\mu} - H_{\mu}(W-m)\right].
\end{align}
It is clear that, because the term inside the square root is real and greater than or equal to zero, for gauge fields which are sufficiently smooth and where $D_W ^{\dagger}D_W$ does not have an exact zero eigenvalue, the Fourier representation of the overlap operator is an analytic function of the momentum, and that the momentum is bound $-\pi\le pa\le \pi$. From the Paley-Weiner theorem~\cite{Paley-Weiner}, this is enough to ensure exponential locality. Numerical experience has shown that the overlap operator remains exponentially local even when $D_W ^{\dagger}D_W$ has a zero eigenvalue~\cite{KriegPersonalCommunication}. $\tilde{D}_O(p)$ is only zero at $p=0$, meaning that it has no unwanted doublers. It is also clear that it has the correct continuum limit. The overlap topological charge is
\begin{gather}
Q_O = -\frac{1}{2}\Tr(\epsilon(\gamma_5 D_W))
\end{gather}

\subsection{The eigenvalue spectrum of generalised Ginsparg-Wilson operators}\label{sec:comment}
I now demonstrate that those chiral Dirac operators satisfying both $[D,D^{\dagger}] = 0$ and the Ginsparg-Wilson equation have an eigenvalue spectrum on a curve in the complex plane. This condition, along with $\gamma_5$-Hermiticity, guarantees that the eigenvalue spectrum is symmetric under reflection in the real axis. 

 I start with the Ginsparg Wilson equation
\begin{gather}
D + D^{\dagger} = D^{\dagger}RD.\label{eq:GW3}
\end{gather}
Using a spectral decomposition, which is valid if $D$ and $D^{\dagger}$ share eigenvalues, i.e. $[D,D^{\dagger}] = 0$, I obtain
\begin{gather}
2 \text{Re}(\lambda) = |\lambda|^2(\psi,R\psi).\label{eq:spectraldecomposition}
\end{gather}
It is clear that equation (\ref{eq:GW3}) implies that
\begin{gather}
[D,D^{\dagger}](1-RD) = D^{\dagger}[D,R]D. 
\end{gather}
Thus, if $[D,D^{\dagger}]=0$ then $[D,R]=0$ and $\psi$ is also an eigenvector of $R$. This enables me to write
\begin{gather}
2 \text{Re}(\lambda) = |\lambda|^2R'(\lambda),
\end{gather}
where $R'$ is some real function of the eigenvalue. Thus, in this case, the eigenvalue spectrum is constrained to a curve in the complex plane.

In the general case, the eigenvalues of a Ginsparg-Wilson operator will not lie on a curve. For example, we can consider a Dirac operator $D$, defined by~\cite{Chiu:1999gq}
\begin{gather}
T^{\dagger}DT = D_{GW},
\end{gather}
where $D_{GW}\in\mathfrak{D}_C$ is some Ginsparg-Wilson operator obeying the relation
\begin{gather}
D_{GW}\gamma_5 + \gamma_5D_{GW} = D_{GW}\gamma_5R_{GW}D_{GW},
\end{gather} 
and $T$ is some local operator which commutes with $\gamma_5$, satisfies the correct continuum limit and whose inverse both exists and is local. In the context of the renormalisation group and the fixed point action, $T$ is equivalent to a change in the blocking procedure used to modify the lattice spacing. Given that $T$ is invertible,
$D$ will satisfy a Ginsparg-Wilson equation
\begin{gather}
D\gamma_5 + \gamma_5 D = D\gamma_5 T R_{GW} T^{\dagger} D.
\end{gather}
For every vector satisfying $D_{GW}\ket{\psi} = 0$ there will be a vector $T^{-1}\ket{\psi}$ which is a zero eigenvector of $D$. Otherwise there will (in general) be no relation between the eigenvalues and eigenvectors of $D$ and $D_{GW}$ unless $T$ commutes with $D_{GW}$ or is unitary. Every particular Ginsparg-Wilson operator is related to at least one operator within the class $\mathfrak{D}_C$ by some transformation $T$. The index of the Dirac operator is unchanged:
\begin{gather}
Q_T = -\frac{1}{2}\Tr\;  R D = -\frac{1}{2}\Tr\; T R_{GW} T^{\dagger} (T^{\dagger})^{-1} D_{GW} (T)^{-1} = Q_{GW}.
\end{gather}
Not every operator within $\mathfrak{D}_C$ can be mapped to every other operator in $\mathfrak{D}_C$. To move from one Dirac operator to another which commutes with it, it is necessary to use a transformation $T$ which commutes with both Dirac operators, and both $R$ functions. In some cases $R$ might have some zero eigenvalues, and the transformation needed to increase the number of zero eigenvalues of $R$ is not invertible.  
\section{Additional solutions to the Ginsparg-Wilson equation}\label{sec:3}
Consider the Dirac operator
\begin{gather}
D_r = 1 + r\left[\frac{1}{2}\big(\gamma_5 \epsilon(\gamma_5 D_W) + \epsilon (\gamma_5 D_W) \gamma_5\big)\right] \gamma_5 \epsilon(\gamma_5 D_W),\label{eq:20}
\end{gather}
where $r[x]$ is some real, positive, and analytic function (which, as I shall demonstrate, is enough to ensure locality), and which satisfies 
\begin{gather}
r[\pm 1] = 1\label{eq:req1}.
\end{gather}
The condition given in equation (\ref{eq:req1}) ensures that, for a suitable choice of $D_W$, such as the one given in equation (\ref{eq:DW}), the operator has the correct continuum limit and no doublers. It is trivial to show that because $r$ commutes with both $\gamma_5$ and $\epsilon$, $D_r$ is $\gamma_5$-Hermitian. Furthermore, it satisfies the Ginsparg-Wilson relation with 
\begin{align}
S = & 1\nonumber\\
R_r =& \frac{1}{D}(2+r\left[\frac{1}{2}\left(\epsilon\gamma_5 + \gamma_5 \epsilon)\right](\epsilon\gamma_5 + \gamma_5 \epsilon)\right) \frac{1}{D^{\dagger}} \nonumber\\=& \frac{2+(\epsilon\gamma_5 + \gamma_5 \epsilon)r\left[\frac{1}{2}(\epsilon\gamma_5 + \gamma_5 \epsilon)\right]}{1 + r\left[\frac{1}{2}(\epsilon\gamma_5 + \gamma_5 \epsilon)\right]^2 + r\left[\frac{1}{2}(\epsilon\gamma_5 + \gamma_5 \epsilon)\right](\epsilon\gamma_5 + \gamma_5 \epsilon)}. 
\end{align}
From equation (\ref{eq:dwp}), I obtain
\begin{align}
\tilde{D}_r(p) =& 1 + r\left[x\right](\gamma_{\mu}H_{\mu} + W - m) \frac{1}{\sqrt{\tilde{B}(p)}}\nonumber\\
x = &\frac{1}{2}\left[(\gamma_\mu H_{\mu}(p) + W(p) - m)\frac{1}{\sqrt{\tilde{B}}(p)} +\frac{1}{\sqrt{\tilde{B}}(p)}(-\gamma_\mu H_{\mu}(p) + W(p) - m)\right] .\label{eq:Drp}
\end{align}
It is clear that if the gauge field $A$ and the function $r$ are analytic and $D_W^{\dagger}D_W$ has no zero eigenvalues, then the Fourier representation of the Dirac operator is analytic. Therefore, using the same argument as for the overlap operator, it is exponentially local. It is also clear, given equation (\ref{eq:req1}), that these Dirac operators have the correct continuum limit and that the doublers have infinite mass. In the momentum representation, $\tilde{R}$ is given by
\begin{gather}
\tilde{R}_r(p) = \frac{2+2r\left[x\right]x}{1 +  (r\left[x\right] )^2 + 2r\left[x\right] x}.
\end{gather}
There is potentially a pole in $\tilde{R}_r$ at $1+r[x]^2 + 2r[x]x = 0 $ or, equivalently, at $r = -x + \sqrt{x^2 - 1}$. Since $r$ and $x$ are both real, $-1\le x \le 1$ and $r$ is positive, the pole cannot contribute for any $x \neq -1$. 
Expanding $\tilde{R}_r$ around $x  = -1$ gives
\begin{gather}
\tilde{R}_r = \frac{2(1-r'[-1])(1+x)+\frac{1}{2}(1+x)^2(2r'[-1] - r''[-1])+O((1+x)^3)}{ 2(1+x) + \frac{1}{2}(1+x)^2(2r'[-1]^2 + 2r'[-1] + r''[-1])+O((1+x)^3)}.
\end{gather}
Therefore, $\tilde{R}_r$ has a smooth limit to $x=-1$ and does not diverge. $\tilde{B}(p)$ is always real and positive (again assuming analytic $r$, $A$ and that $D_W^{\dagger}D_W$ has no exact zero eigenvalues), so $\tilde{R}_r$ is analytic, and $R_r$ will (at worst) fall of exponentially with distance. Thus both the Dirac operator and the Ginsparg-Wilson function $R_r$ are exponentially local, and this operator should be a suitable lattice Dirac operator with chiral symmetry. Given my experience with the overlap operator, I do not expect locality to break down when $D_W^{\dagger}D_W$ has an exact zero eigenvalue.

I now need to demonstrate that these operators have a well defined topological charge; which is most easily done by comparing the eigenvalue spectrum of this operator with that of the overlap operator. I write the non-zero modes of the Hermitian overlap operator as $\ket{\psi_{+}}$ and $\ket{\psi_{-}}$, where the zero modes are $\ket{\psi_0}$ and the unpaired eigenvectors of the overlap operator with eigenvalue $\pm2$ are $\ket{\psi_2}$.
Since $D_O$ and $D_r$ commute, it is easy to show,  that these eigenvectors are also eigenvectors of $\gamma_5 D_r$ and $R_r$. Additionally, it can be shown that
\begin{align}
D_r\ket{\psi_0} = & 0\nonumber\\
\bra{\psi_2}D_r\ket{\psi_2} = & \bra{\psi_2}D_O\ket{\psi_2}\nonumber\\
\bra{\psi_+}D_r\ket{\psi_+}=& - \bra{\psi_-}D_r\ket{\psi_-}\nonumber\\
R_r \ket{\psi_0} =& R_r \ket{\psi_2} = 1\nonumber\\
R_r\ket{\psi_+} = &R_r \ket{\psi_-}.
\end{align}
Hence,
\begin{gather}
-\frac{1}{2}\Tr\;\gamma_5R_rD_r = -\frac{1}{2}\Tr\;\gamma_5D_O = Q_f.
\end{gather}
Therefore these Dirac operators will have the same zero modes and topological charge as the overlap operator.

I note that writing the operator as $r(\cos\theta) e^{i\theta}$, which I have done in equation (\ref{eq:20}), is only possible because $[D_O,D^{\dagger}_O] = 0$. It would not be possible to perform a similar decomposition with, for example, the Wilson operator to shift the Wilson eigenvalues to a closed curve on the complex plane.
\section{Numerical tests}\label{sec:4}
To test the locality of the Dirac operator $D_r$ and the Ginsparg-Wilson function $R$, I measured 
\begin{align}
L_D(x,y) =& \langle\phi(x)|D_r|\phi(y)\rangle,\nonumber\\
L_R(x,y) =& \langle\phi(x)|R_r|\phi(y)\rangle
\end{align}
where $\phi(x)$ is a point source. I then plot the mean value of the locality $L(|x-y| = d)$ against $d$, the distance between $x$ and $y$ in lattice units.

\begin{figure}
\begin{center}
\begin{tabular}{c}
\includegraphics[width = 13.5cm]{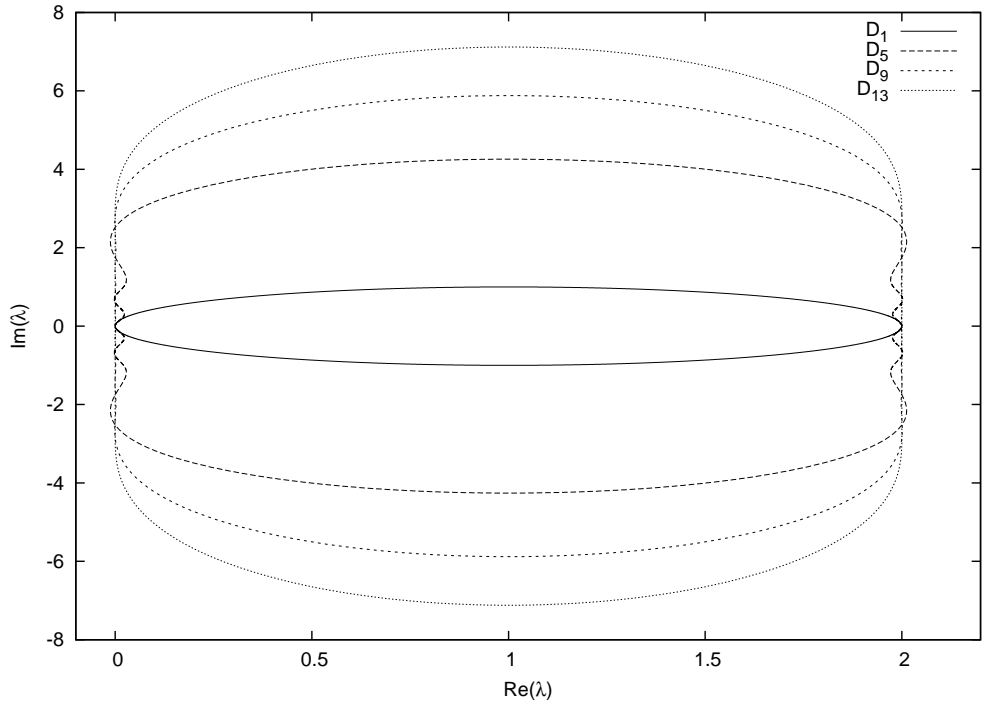}
\end{tabular}
\caption{The eigenvalue distributions of the various Dirac operators used in this study.}\label{fig:1}
\end{center}
\end{figure}

\begin{figure}
\begin{center}
\begin{tabular}{c}
\includegraphics[width = 13.5cm]{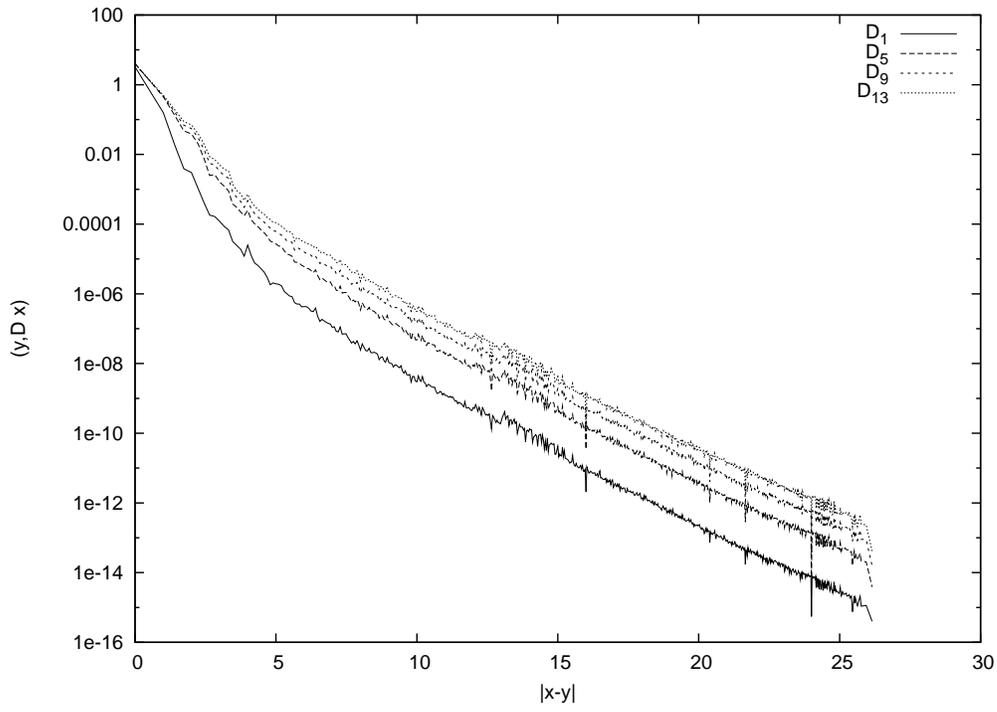}
\end{tabular}
\caption{The locality function $L_D$ on a $12^348$ dynamical overlap ensemble.}\label{fig:firstfig}
\end{center}
\end{figure}

\begin{figure}
\begin{center}
\begin{tabular}{c}
\includegraphics[width = 13.5cm]{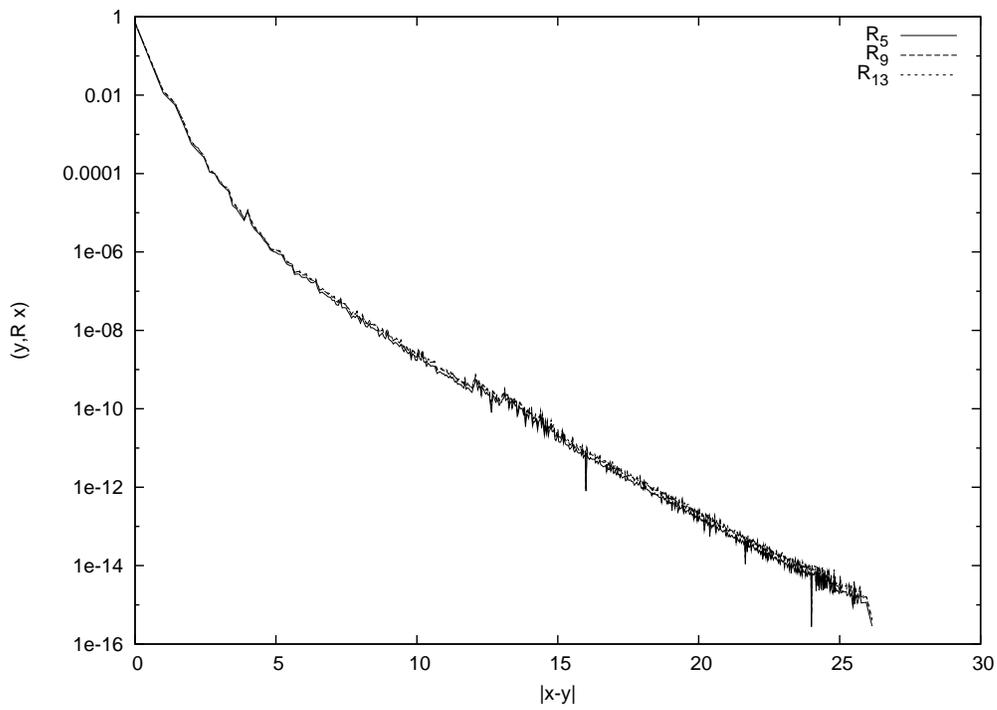}
\end{tabular}
\caption{The locality function $L_R$ on a $12^348$ dynamical overlap ensemble.}\label{fig:lastfig}
\end{center}
\end{figure}

The operator with $r=1/|x|$ mimics the continuum operator in that the eigenvalue spectrum is on two lines --- one along the imaginary axis containing the physical modes, and one at a constant mass (in the continuum limit an infinite mass), containing the doublers. However, because this operator is impractical,\footnote{Since $H_{\mu}$ and $W$ do not commute in the non-Abelian theory, I have only been to write the operator in a simplified form in the free or Abelian theory. To calculate the operator exactly would require calculating eigenvalues of the overlap operator to a high precision and using some rational or polynomial approximation to $|x|$ to simulate the rest of the eigenvalue spectrum, which is possible but impractical. It is also not clear if this operator is local: $r$ is not analytic; however we can construct successively better approximations to $r$ all of which will be local. I also note that because $R_{1/|x|} = 0$ for the physical modes (it is not zero for the eigenvalues corresponding to the doublers; so that the Nielson-Ninomiya theorem is still avoided) the left and right chiral projectors are identical for the physical modes. Thus if this operator is local, it might be possible to use it to circumvent the CP and T violations described in~\cite{Hasenfratz:2005ch}. However, a full discussion of this interesting topic is beyond the scope of this paper, and deserves a full treatment in a future work.} I shall use an approximation to this function. I therefore set $r$ as the Chebyshev approximations (over the range $0.1<x<1$) of order $n-1$ of the function $f = 1/\sqrt{x^2}$, for $n = 1,5,9$ and $13$, normalised so that it satisfies equation (\ref{eq:req1}). This gives me four different Dirac operators, $D_1$, $D_5$, $D_9$ and $D_{13}$. The (theoretical) eigenvalue distributions for the operators $D_n$ are shown in figure \ref{fig:1}. $D_1$ is the overlap operator. The kernel of the matrix sign function, $D_W$, is as defined by equation (\ref{eq:DW}), but with two levels of stout smearing~\cite{Morningstar:2003gk} at parameter 0.1, and with $m = 1.5$.
I tested the locality of these operators on configurations from a $12^348$ Dynamical overlap ensemble, with lattice spacing $a\sim \text{0.13}\text{fm}$. I apply the Dirac operator $D_n$ or Ginsparg-Wilson function $R_n$ to a unit source at $x$, and calculate its projection onto another unit source vector at $y$, averaging over $x$, $y$ and configurations. The results are shown in figures \ref{fig:firstfig} and \ref{fig:lastfig}. 


It can be seen that in all cases both the Dirac operator and the Ginsparg-Wilson function are exponentially local, with the same rate of decay as the overlap operator. They are all constrained so that $L_D < \alpha_D e^{-\beta|x-y|}$ and $L_R < \alpha_R e^{-\beta|x-y|}$. The value of $\beta$ seems to be independent of $n$, and (we might conjecture) seems to be general across all possible functions $r$. The value of $\alpha$ increases with increasing $n$ for the Dirac operator (although there is very small change for the Ginsparg-Wilson function), and does depend on $r$. It is possible to suspect that other forms of the Ginsparg-Wilson Dirac operator could have improved locality compared to the overlap operator.\footnote{For the same kernel. It is known that using an improved kernel can improve the locality~\cite{Bietenholz:2002ks}, but this is a separate issue.} It can be concluded that these Dirac operators and Ginsparg Wilson functions are, as expected, exponentially local. The various small wiggles in the curves are caused by the geometry of the lattice and the breaking of rotational symmetry.

\section{Generalisation of the Ginsparg-Wilson Dirac operator}\label{sec:5}

The proposed operator given in equation (\ref{eq:20}) can be easily generalised further. One possibility is to multiply by a function $q$:
\begin{gather}
D_{rq} = q\left[\frac{1}{2}\left(\gamma_5\epsilon + \epsilon \gamma_5\right)\right]\left(1 + r \left[\frac{1}{2}(\gamma_5\epsilon + \epsilon \gamma_5)\right]\gamma_5\epsilon\right),
\end{gather}
where we again have the constraints that $q[x]$ must be positive, analytic, and $q[1] = q[-1]$ (for the sake of being definite, I shall use $q[\pm1] = 1$, although the value can easily be absorbed into the fermion renormalisation constant), and, to give the correct continuum limit, restricted to $q = 1 + O(a)$. This leads to a Ginsparg-Wilson function
\begin{gather}
R_{rq} = \frac{1}{q[x]} R_r[x],
\end{gather}
and it is clear that, with $q$ as specified, this contains no poles in the Fourier representation, so $R_{rq}$ will be local. We can, of course, generalise further, and consider an operator
\begin{gather}
D_{rqth} = t\left[\frac{1}{2}(\gamma_5\epsilon + \epsilon \gamma_5)\right]h\left(q\left[\frac{1}{2}(\gamma_5\epsilon + \epsilon \gamma_5)\right]\left(1 + r \left[\frac{1}{2}(\gamma_5\epsilon + \epsilon \gamma_5)\right]\gamma_5\epsilon\right)\right),\label{eq:Drsth}
\end{gather} 
where $h(z)$ is analytic, positive except for $z = 0$ and satisfies $h(0) = 0$, $h(2) > 0 $ and $h(z)^{\dagger} = h(z^{\dagger})$; and $t$ obeys the same conditions as $q$ and $r$. Finally, I can choose $S\neq 1$ in equation (\ref{eq:GW}), which will modify the above Dirac operator to give:
\begin{gather}
D_{rqthS} = \frac{1}{S(\gamma_5\epsilon)^{\dagger}}t\left[\frac{1}{2}(\gamma_5\epsilon + \epsilon \gamma_5)\right]h\left(q\left[\frac{1}{2}(\gamma_5\epsilon + \epsilon \gamma_5)\right]\left(1 + r \left[\frac{1}{2}(\gamma_5\epsilon + \epsilon \gamma_5)\right]\gamma_5\epsilon\right)\right),\label{eq:Drqths}
\end{gather}
where $S(z)$ is an analytic function satisfying $S(-1) = S(1) = 1$ and $S(z^{\dagger}) = S(z)^{\dagger}$.
Once again, from the Fourier representation, it can be shown that $D_{rqthS}$ and $R_{rqthS}$ are both exponentially local and have the correct continuum limit. Fujikawa's operators~\cite{Fujikawa:2000my,*Fujikawa:2001fb} are members of this class, with $r=q=S=t= 1$, $h(z) = z^{1/(2k+1)}$, and a particular form for the the kernel operator $D_W$ which gives the correct continuum limit. 

Kerler~\cite{Kerler:2002xk,*Kerler:2002fi,*Kerler:2003qv} considered, at first, a alternative form of lattice chiral symmetry
\begin{gather}
\gamma_5 D + D \gamma_5 V = 0, \label{eq:Kerler}
\end{gather}
where $V$ is unitary and $\gamma_5$-Hermitian. This relates to the standard Ginsparg-Wilson relation with the substitution of variables $V = 1-RD = -\frac{1}{D^{\dagger}}D$. He later extended equation (\ref{eq:Kerler}) to include some possible $S\neq 1$ forms of the Ginsparg-Wilson equation (following the transformations outlined in ~\cite{Hasenfratz:2005ch}), and, by considering Dirac operators which are functions of $V$ and using a spectral decomposition, he
 showed that operators of the form
\begin{gather}
D_K = -i(\overline{G}(V) G(V))^{1/2}H\left(\frac{1}{2i}(V^{-1/2} - V^{1/2})W\left[\frac{V + V^{\dagger}}{2}\right]\right),
\end{gather}
where $H(-z) = -H[z]$ (with the expansion around $z=0$ giving the correct continuum limit; for example using $D_W$ as the kernel of the matrix sign sign function, $H[z] = z +O(z^2)$), $W(-1) \neq 0$, both functions are Hermitian and 
\begin{align}
G(V) = &((1-s_k) + s_kV)/N\nonumber\\
\overline{G} = & (s_k + (1-s_k)V)/N\nonumber\\
N = &\sqrt{1-2s_k(1-s_k)(1-\frac{1}{2}(V + V^{\dagger})},
\end{align}
satisfy a lattice chiral symmetry. It is clear that Kerler's solution (with $V = \gamma_5 \epsilon$) and my own have a certain similarity, although they were derived from different approaches: his from attempting an algebraic solution of the eigenvalue equivalent of the Ginsparg-Wilson equation; mine from desiring to test a seemingly general property of chiral Dirac operators. His solutions are a particular form of my most general $D_{rqthS}$ solution, specifically with $r = 1$ and with a particular form of $S$ and $t$. Kerler did not discuss in detail the locality of the Ginsparg-Wilson function, which is crucial in determining if we truly have a lattice chiral symmetry. It is unclear whether these generalisations offer any advantage over the operator given in the previous sections.\footnote{One possibility might be to absorb a perturbative expansion of the fermion renormalisation into the Dirac operator; or to otherwise reduce the higher order lattice artifacts.}

Finally, it is worth spending a moment considering whether it is necessary to use $\gamma_5 \epsilon$ to construct these operators, and not some other unitary operator $u$. There are several ways in which we can construct $u$:
\begin{enumerate}
\item $u = C/\sqrt{C^{\dagger}C}$; 
\item $u = \frac{1}{C^{\dagger} }C$ for $C$ not (anti-)Hermitian;
\item $u = e^{A}$, for anti-Hermitian $A$.
\end{enumerate} 
$u$, of course, would have to be local and have the correct continuum limit. I have already discussed the first option. For the second option, forcing $u$ to be local, having the correct continuum limit and being $\gamma_5$-Hermitian will place considerable restrictions on the possible choices of $C$ (one option is $C = 1 + D_{GW}$, where $D_{GW}$ is another Ginsparg-Wilson operator; but if we choose to use a Ginsparg-Wilson operator constructed the matrix sign function this will revert to another form of equation (\ref{eq:Drqths}). For the third option, it is not obvious how to construct suitable operators which are both $\gamma_5$-Hermitian and free of doublers; for example $i\gamma_5(1-e^{i\pi/|\beta|\gamma_5D_W})$ is not $\gamma_5$-Hermitian, and $1-e^{D - \gamma_5D\gamma_5}$ has doublers. Thus using the matrix sign function remains the only currently known possibility to construct local chiral lattice Dirac operators free from doublers.

One can also, of course, transform the Dirac operators using the method outlined in section \ref{sec:comment}. By choosing $T$ and $T^{\dagger}$ which commute with $D$, we can map the eigenvalue spectrum to any closed curve (which passes through the origin to get the correct continuum limit and through two to remove the doublers). The only restriction with the mapping is caused by the number of zero modes of $R$; but since $R_{rqthS}$ has an arbitrary number of zero modes depending on the choice of functions, we can map to any possible $R$ and hence $D$. Thus the eigenvalue spectrum of any Dirac operator within $\mathfrak{D}_C$ can be mapped to the eigenvalue spectrum of at least one of the operators $D_{rqthS}$ by a suitable transformation. The eigenvectors of $D_{rqthS}$ are determined by $D_{W}$, and can be freely modified (within certain constraints) by adjusting $D_{W}$. Given that the Dirac operators can be defined in terms of their eigenvalues and eigenvectors spectrum, it is plausible, if not yet proven, that all Dirac operators within $\mathfrak{D}_C$ are Ginsparg-Wilson operators.

\section{Conclusion}\label{sec:conclusion}
I have demonstrated that the overlap operator is just the simplest member of a class of chiral Dirac operators by constructing additional Ginsparg-Wilson operators. I propose that any lattice Dirac operator which has
\begin{enumerate}
\item The correct continuum limit with no doublers;
\item Eigenvalues which lie on a closed loop in the complex plane, symmetric under reflection in the imaginary axis, and single valued with respect to the angle from the center of the Ginsparg-Wilson circle;
\end{enumerate}
will be exponentially local (in a sufficiently large volume and sufficiently smooth gauge fields), will satisfy a lattice chiral symmetry with exponentially local (or better) Ginsparg-Wilson function $R$ with an exact index theorem, will be $\gamma_5$-Hermitian, and will thus be a suitable lattice Dirac operator. Of course, these additional Dirac operators are more expensive to simulate than the overlap operator while it is not clear that they have any benefits over the overlap operator. Thus it is unlikely that they will have any more than theoretical interest.

Furthermore, I have shown that chiral Dirac operators fall into certain homotopy classes determined by the number of zero modes of the Ginsparg-Wilson function $R$. In the continuum limit, all these Dirac operators will reduce to the same, universal, operator, and it is most unlikely that this division has any physical relevance. However, it seems likely that the perfect action (the lattice action with no discretization errors) will have to fall into the same homotopy class as the continuum Dirac operator. 
\section*{Acknowledgements}
I am grateful of support from the DFG. Computations were carried out on a Linux cluster at the J\"ulich Supercomputing center. I am grateful for useful discussions with Andreas Sch\"afer, Stephan Durr and Wolfgang Bietenholz, and the referee for pointing out an error in the first draft of this paper.
\bibliographystyle{modified_cpc}
\bibliography{new_GW_operator}

\begin{thebibliography}{10}

\bibitem{Nishy-Ninny}
Nielsen, H.~B. and Ninomiya, B.,
\newblock Nucl. Phys {\bf B185} (1981) 20.

\bibitem{Ginsparg:1982bj}
Ginsparg, P.~H. and Wilson, K.~G.,
\newblock Phys. Rev. {\bf D25} (1982) 2649.

\bibitem{Hasenfratz:1998ri}
Hasenfratz, P., Laliena, V., and Niedermayer, F.,
\newblock Phys. Lett. {\bf B427} (1998) 125,
\newblock hep-lat/9801021.

\bibitem{Luscher:1998pqa}
L{\"u}scher, M.,
\newblock Phys. Lett. {\bf B428} (1998) 342,
\newblock hep-lat/9802011.

\bibitem{Kaplan:1992bt}
Kaplan, D.~B.,
\newblock Phys. Lett. {\bf B288} (1992) 342,
\newblock hep-lat/9206013.

\bibitem{Narayanan:1993sk}
Narayanan, R. and Neuberger, H.,
\newblock Nucl. Phys. {\bf B412} (1994) 574,
\newblock hep-lat/9307006.

\bibitem{Narayanan:1993ss}
Narayanan, R. and Neuberger, H.,
\newblock Phys. Rev. Lett. {\bf 71} (1993) 3251,
\newblock hep-lat/9308011.

\bibitem{deA.Bicudo:1999wq}
de~A.~Bicudo, P.~J.,
\newblock Phys. Lett. {\bf B478} (2000) 379,
\newblock hep-lat/9909157.

\bibitem{Hasenfratz:1994}
Hasenfratz, P. and Niedermayer, F.,
\newblock Nucl. Phys. {\bf B414} (1994) 785.

\bibitem{Bietenholz:1995nk}
Bietenholz, W. and Wiese, U.~J.,
\newblock Phys. Lett. {\bf B378} (1996) 222,
\newblock hep-lat/9503022.

\bibitem{Fujikawa:2000my}
Fujikawa, K.,
\newblock Nucl. Phys. {\bf B589} (2000) 487,
\newblock hep-lat/0004012.

\bibitem{Fujikawa:2001fb}
Fujikawa, K. and Ishibashi, M.,
\newblock Nucl. Phys. Proc. Suppl. {\bf 106} (2002) 712,
\newblock hep-lat/0110023.

\bibitem{Kerler:2002xk}
Kerler, W.,
\newblock Nucl. Phys. {\bf B646} (2002) 201,
\newblock hep-lat/0204008.

\bibitem{Kerler:2002fi}
Kerler, W.,
\newblock Int. J. Mod. Phys. {\bf A18} (2003) 2565,
\newblock hep-lat/0212021.

\bibitem{Kerler:2003qv}
Kerler, W.,
\newblock Nucl. Phys. {\bf B680} (2004) 51,
\newblock hep-lat/0307011.

\bibitem{Hasenfratz:1998jp}
Hasenfratz, P.,
\newblock Nucl. Phys. {\bf B525} (1998) 401,
\newblock hep-lat/9802007.

\bibitem{Horvath:1999bk}
Horvath, I.,
\newblock Phys. Rev. {\bf D60} (1999) 034510,
\newblock hep-lat/9901014.

\bibitem{Hernandez:1998et}
Hernandez, P., Jansen, K., and Luscher, M.,
\newblock Nucl. Phys. {\bf B552} (1999) 363,
\newblock hep-lat/9808010.

\bibitem{BietenholzPersonalCommunication}
I thank Wolfgang Bietenholz for bringing the relevant section of this paper to
  my attention.

\bibitem{Kikukawa:1998pd}
Kikukawa, Y. and Yamada, A.,
\newblock Phys. Lett. {\bf B448} (1999) 265,
\newblock hep-lat/9806013.

\bibitem{Adams:1998eg}
Adams, D.~H.,
\newblock Annals Phys. {\bf 296} (2002) 131,
\newblock hep-lat/9812003.

\bibitem{Horvath:2006md}
Horvath, I.,
\newblock (2006),
\newblock hep-lat/0607031.

\bibitem{Liu:2007hq}
Liu, K.~F., Alexandru, A., and Horvath, I.,
\newblock (2007),
\newblock hep-lat/0703010.

\bibitem{Bietenholz:1999dh}
Bietenholz, W.,
\newblock (1999),
\newblock hep-lat/0001001.

\bibitem{Neuberger:1998my}
Neuberger, H.,
\newblock Phys. Rev. Lett. {\bf 81} (1998) 4060,
\newblock hep-lat/9806025.

\bibitem{Paley-Weiner}
Paley, R. and Wiener, N.,
\newblock {\em Fourier Tansform in Complex Domain},
\newblock Providence, R. I., 1934,
\newblock Theorem XII.

\bibitem{KriegPersonalCommunication}
Krieg, S.,
\newblock Internal collaboration report.

\bibitem{Chiu:1999gq}
Chiu, T.-W.,
\newblock Phys. Lett. {\bf B474} (2000) 89,
\newblock hep-lat/9910029.

\bibitem{Hasenfratz:2005ch}
Hasenfratz, P. and Bissegger, M.,
\newblock Phys. Lett. {\bf B613} (2005) 57,
\newblock hep-lat/0501010.

\bibitem{Morningstar:2003gk}
Morningstar, C. and Peardon, M.~J.,
\newblock Phys. Rev. {\bf D69} (2004) 054501,
\newblock hep-lat/0311018.

\bibitem{Bietenholz:2002ks}
Bietenholz, W.,
\newblock Nucl. Phys. {\bf B644} (2002) 223,
\newblock hep-lat/0204016.

\end{thebibliography}

\end{document}